\documentclass[final,3p,times,twocolumn]{elsarticle}
\usepackage{amssymb}
\usepackage{soul}
\usepackage{amsmath}
\usepackage{float}
\usepackage{xcolor}
\definecolor{RED}{RGB}{156,78,90}

\journal{Chemical Engineering Journal}

\usepackage{lineno}

\usepackage{color}
\definecolor{B}{RGB}{52,78,200}

\begin{document}
\begin{sloppypar}
\begin{frontmatter}

\title{Uncovering the nanoscopic phase behavior of ternary solutions in the presence of electrolytes: from pre-Ouzo to Ouzo region}

\author[1]{Mingbo Li\corref{cor1}}
\ead{mingboli@sjtu.edu.cn}
\cortext[cor1]{Corresponding author}
\address[1]{Key Laboratory of Hydrodynamics (Ministry of Education), School of Ocean and Civil Engineering, Shanghai Jiao Tong University, Shanghai 200240, China}

\author[2]{Rushi Lai}
\address[2]{New Cornerstone Science Laboratory, Center for Combustion Energy, Key Laboratory for Thermal Science and Power Engineering of MoE, Department of Energy and Power Engineering, Tsinghua University, Beijing 100084, China}

\author[1]{Yadi Tian}

\author[2]{Yawen Gao\corref{cor1}}
\ead{yawengao@mail.tsinghua.edu.cn}

\author[1]{Benlong Wang}

\author[2,3]{Chao Sun}
\address[3]{Department of Engineering Mechanics, School of Aerospace Engineering, Tsinghua University, Beijing 100084, China}

\begin{abstract}
In this work, we report a comprehensive study of how electrolyte addition governs the structure and stability of surfactant-free microemulsions in a trans-anethol/ethanol/water system. The universal structural response has been validated, spanning the full range of solution dispersed-phase structurings, from sub-10 nm W/O reverse-aggregates to O/W mesoscopic droplets ($\sim$100 nm) and classical Ouzo droplets ($\sim$1 $\mu$m). Experimental results reveal that there is a threshold for electrolyte levels above which oil-in-water nanodroplet coalescence and phase separation are triggered: screening of electrical double layers and "salting out" of hydrophobic components drives hydrotrope into fewer, larger droplets. The total oil volume sequestered in the dispersed phase remains essentially constant, indicating oil redistribution rather than dissolution. In contrast, water-in-oil nanodroplets in a predominantly organic medium display near-complete insensitivity to ionic strength, owing to low dielectric screening and tight interfacial packing that exclude substantial ion uptake. Finally, addition of high salt to Ouzo droplets accelerates their collapse: large droplets fuse and sediment, leaving only residual nanostructures and confirming electrolyte-driven phase demixing. This insight offers clear guidelines for designing additive-free emulsions with tailored lifetimes and nanostructure architectures across pharmaceutical, food, and materials applications.  
\end{abstract}


\end{frontmatter}


\section{Introduction}

Surfactant‐free microemulsions (SFMEs) are thermodynamically metastable dispersions of one liquid phase within another, stabilized not by conventional surfactants but by the intrinsic mutual solubility of the components and subtle interfacial forces~\cite{subramanian2013mesoscale, schoettl2014emergence, hou2016surfactant, rak2018mesoscale, li2022spontaneously}. Unlike classical microemulsions, which require amphiphilic molecules to lower interfacial tension and kinetically arrest droplet coalescence, SFMEs exploit cosolvents (also named as "amphi-solvent" or "hydrotrope") to mediate oil–water miscibility and form nano-domains without added surfactant. These pseudophase nanostructurings, which form spontaneously, are different from "standard" swollen micelles and are given various names, e.g., "pre-Ouzo"~\cite{klossek2012structure}, micellar-like structural fluctuations~\cite{novikov2017dual}, mesoscale inhomogeneities~\cite{hahn2019ab}, and so on~\cite{smith1977oil, prevost2016small, subramanian2013mesoscale, robertson2016mesoscale}. Since the seminal reports of spontaneously forming "Ouzo" emulsions~\cite{vitale2003liquid, sitnikova2005spontaneously}, where anethol precipitates as droplets upon dilution of anethol/ethanol/water solutions, interest in SFME systems has grown rapidly due to their simplicity and potential for "clean" formulations. 

In parallel, fundamental studies have revealed multiscale nanostructuring in various SFME systems, from molecular-scale aggregates ($\sim$1 nm)~\cite{schottl2019combined, diat2013octanol, ashfaq2024uncovering} to mesoscopic droplets ($\sim$100 nm)~\cite{li2022spontaneously, rak2018mesoscale}. The origin and thermodynamic stability of these well-organized nano-domains have been a hot topic. The hydrophobic nature of the oil-phase can drive aggregation, and the stability of these nanophases has been attributed to the balance between hydration forces and entropy~\cite{zemb2016explain, ma2025thermodynamic}. Depending on the composition of mixture, it is suggested that different morphologies, like oil-in-water (O/W), water-in-oil (W/O), and bicontinuous structures can result. The cosolvent is unable to self-assemble into micelles or form organized films at the oil–water interface. However, they can assist in the formation of dynamic molecular clusters containing both oil and cosolvent through interactions like hydrogen bonding, and may be enriched specifically at the interface between the oil-rich and water-rich regions~\cite{lund1980detergentless, schoettl2014emergence, diat2013octanol}. Its behavior resembles more the weak specific attraction of ions to interfaces~\cite{jungwirth2006specific}. Molecular dynamics simulation results confirm this and show that the ethanol molecules enriched at the interface further promote hydration interactions~\cite{marcelja1997hydration, marvcelja2011hydration, donaldson2015developing, lopian2016morphologies, chen2024unraveling}. From other viewpoints, hydrogen bonds and $n-\Pi^*$ interactions between the oil and cosolvent strongly support the development of the nanophases, especially in the volumetric structures containing a smaller amount of oil and a larger amount of water~\cite{ashfaq2024uncovering}. A variety of molecular self-assembly pathways have been observed in nucleating solutions, emphasising the importance of pre-nucleation microstructures in nucleation for a broad spectrum of organic and inorganic systems~\cite{gebauer2008stable, davey2013nucleation, jawor2015effect}.   

Controlling SFME properties, droplet size, stability, morphology, and lifetime, is essential for their effective exploitation in applications such as controlled drug delivery, emulsion polymerization, and encapsulation of flavors or bioactive compounds. Achieving reproducible droplet formation and longevity in SFME thus demands a detailed understanding of the factors that govern nucleation, growth, and coalescence in the absence of traditional emulsifiers. A variety of physicochemical parameters influence SFME behavior~\cite{jing2022ph, wu2023co2}. Composition defines the phase boundaries at which distinct nanophases emerge, from single-phase molecular solutions to metastable mesophases and finally to demixed two-phase regions~\cite{blahnik2022nanoscopic}. Temperature also plays a role by altering solubility limits and interfacial tension~\cite{li2023thermal}, while pressure can modulate volumetric interactions in high-pressure microemulsions. Cosolvent identity (e.g., ethanol vs. propanol) affects both hydrogen-bonding networks and hydrotropic strength, thereby tuning droplet size distributions. 

One important factor influencing the behavior and stability of SFMEs is the addition of inorganic electrolytes. The presence of electrolytes can significantly affect the physicochemical properties of the microemulsion. In conventional emulsions, added salt can "salt out" hydrophobic species, compress electrical double layers, and modify interfacial charge, leading to coalescence or changes in droplet size according to Derjaguin–Landau–Verwey–Overbeek (DLVO) theory~\cite{marcus2015influence, rambhau1992influence}. Yet in SFME systems, where no surfactant is present to generate a stable charge barrier, the impact of ionic strength on nanostructure formation, droplet stability, and phase behavior remains largely unexplored. 

Existing work on salt effects in SFME has been limited. Some studies have reported that trace impurities or buffering salts can subtly alter the onset of turbidity in Ouzo systems~\cite{ottosson2010influence}, while molecular dynamics simulations suggest that salt ions can drive co-solvent toward oil–water interfaces, reinforcing salting-out of hydrophobic solutes~\cite{schottl2018salt}. However, these insights have not been systematically validated by experiments spanning the full range of SFME nanophases (from sub-10 nm W/O aggregates to classical Ouzo droplets and mesoscopic O/W droplets). In particular, the question of whether added salt simply shifts composition-phase boundaries or actively reorganizes existing oil-rich domains into new size regimes has not been answered.

To address this gap, the present study employs a suite of complementary techniques to map the structural response of SFME dispersion phase to varying electrolyte concentrations. We investigate three representative structurings: (i) an O/W nanophase region featuring mesoscopic droplets ($\sim$100 nm), (ii) a W/O nanophase region with sub-10 nm reverse aggregates, and (iii) the classical Ouzo region of micrometer-scale droplets. These findings not only deepen our fundamental understanding of SFME phase behavior but also provide practical guidelines for designing surfactant-free emulsions with tailored droplet sizes and lifetimes in various applications.

\section{Experimental Section}

\subsection{Materials ans ternary solution preparation} 

The ternary system comprised ultrapure water, ethanol, and trans-anethol, mirroring the classical Ouzo formulation~\cite{lohse2020physicochemical}. Water (initial pH 6.5, 18.2 $\rm{M\Omega\cdot{cm}}$ resistivity) was produced by a Milli-Q purification system (Merck, Germany). Ethanol ($ 99.9\%$ purity) was obtained from Beijing J$\&$K Co. Ltd., and trans-anethol ($>99.0\%$ purity, $M_{\mathrm w}=148.20\ \mathrm{g/mol}$) was purchased from Sigma-Aldrich (Germany). The density, dynamic viscosity, and refractive index of each component are summarized in Table S1 (Supplementary Material). All reagents were of analytical grade and used as received.

\begin{figure}[!t]
\centering
\includegraphics[scale = 0.8]{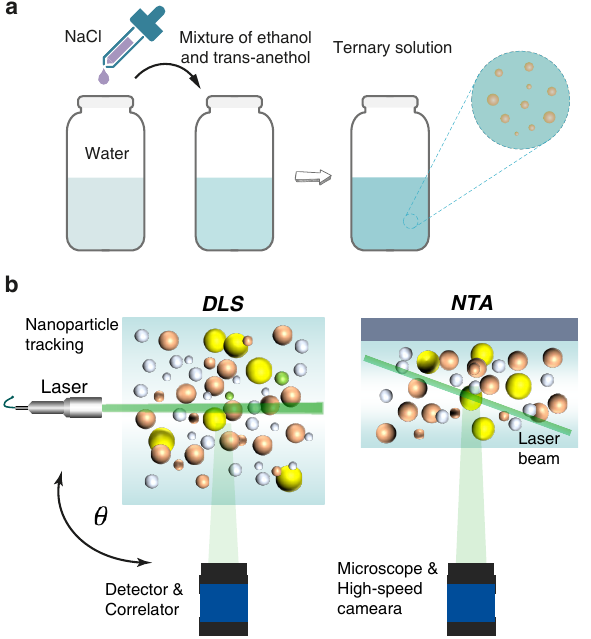}
\caption{(a) Schematic diagram for configuring ternary solutions containing certain concentrations of electrolytes. (b) Schematics of DLS and NTA systems In DLS, one can observe the scattered light of an ensemble of nanoparticles undergoing Brownian motion. In NTA, scattered light of individual nanoparticles is tracked over a period of time.} 
\label{FIG1}
\end{figure}

SFME samples were prepared as follows (see Figure~\ref{FIG1}(a)). Aqueous NaCl solutions spanning $10^{-6}$ to 2.0 M (NaCl$\geq$99.999$\%$, Thermo Scientific) were first formulated. trans-anethol and ethanol were combined at volume fractions $\phi_{o}$ and $\phi_{e}$, respectively, to form a clear, single-phase mixture. The NaCl solution (volume fraction $\phi_{w}$) was then added dropwise under gentle stirring to minimize turbulent mixing. Finally, each sample was vortexed (Vortex-5, Kylin-Bell, China) for at least 3 min to ensure homogeneity and generate a stable SFME.

\subsection{Dynamic Light Scattering (DLS) Measurements }

DLS measurements were performed on a ZEN3700 Zetasizer NanoZSE (Malvern, UK) at a fixed scattering angle of $\theta = 173^\circ$ (schematic in Figure~\ref{FIG1}(b))). Samples ($\sim$1 mL) were loaded into a sealed quartz cell (10 mm light path) to minimize evaporation, then equilibrated for 5 min at 25$^\circ$C prior to measurement. Under these conditions, the instrument resolves hydrodynamic diameters from 0.4 nm to 10 $\mu$m and yields both the intensity autocorrelation functions and the derived size distributions of the nanoscale entities in the bulk. For each condition, two independent measurements were acquired, each comprising at least four repeats to improve statistical reliability. Photon‐counting detectors recorded the scattered‐light intensity, which was then corrected for laser transmittance through the attenuator to calculate the total scattering intensity rate (TSIR, in kcps).
 
\subsection{Nanoparticle Tracking Analysis (NTA) Measurements }

We constructed a custom nanoparticle tracking analysis (NTA) setup to monitor multiscale nanodomains~\cite{siedentopf1902uber, ma2022measurement}. Our instrument comprises three main modules: (i) a laser illumination system, (ii) a detachable microfluidic sample cell, and (iii) a dark‐field imaging system integrating an inverted optical microscope and a high‐speed camera. The schematic diagram is shown in Figure~\ref{FIG1}(b). The sample cell consists of two quartz‐glass plates separated by a fluoroelastomer spacer. The lower plate defines the sample volume, while the thicker upper plate features precision prisms that introduce the laser and solution from opposite sides. Prior to entering the chamber, the well‐collimated laser beam (max. power 150 mW, wavelength 520 nm) undergoes two mirror reflections and two refractions through the prism‐edged optical flat. The sample chamber itself is hexagonal, 500 $\mu$m high, with a maximum diameter of 20 mm.

In the imaging module, a $20\times$ long‐working‐distance objective on an inverted microscope (IX73, Olympus) provides the necessary magnification, while a high‐speed CCD camera (xiD, XIMEA) records dark‐field videos at 30 fps and a spatial resolution of $\sim$227 nm/pixel. For each condition, we capture four 60 s videos. Post‐acquisition, we analyze these sequences in ImageJ/Fiji\cite{schindelin2012fiji} using the NanoTrackJ plugin~\cite{wagner2014dark} to extract Brownian‐particle number concentrations and size distributions. For more details about this NTA, please refer to our previous work~\cite{ma2022measurement}. 

\subsection{Zeta potential, refractive index, conductivity, permittivity and viscosity measurements} 

The Zeta potential of the nanostructures was measured on the same DLS instrument (Zetasizer NanoZSE, Malvern Instruments, UK) using a U-shaped capillary cell. Each sample was analyzed six times at ambient pressure and 25$^\circ$C, and the mean Zeta potential or distribution was reported.

Refractive index measurements were performed immediately after sample preparation with a GR30 refractometer (Shanghai Zhuoguang Instrument Technology, China). The instrument was calibrated at 25$^\circ$C using deionized water before each session, and samples were measured once the refractometer reached the set temperature. To ensure precision, each reading was obtained in triplicate.

Conductivity measurements were performed on a conductometer (METTLER TOLEDO, Switzerland) equipped with a 1 $\rm{cm^{-1}}$ conductivity cell. Prior to analysis, the instrument was calibrated at 25$^\circ$C. Electrolyte samples ($\sim$10 mL) were poured into the cell and covered to minimize evaporation. After temperature equilibration and signal stabilization, the conductivity value was recorded. Each sample was measured in triplicate, with the conductivity averaged over three successive readings. Between measurements, the cell was rinsed with deionized water and blotted dry to prevent cross-contamination.

The dielectric permittivity and dynamic viscosity of the ternary solutions—key parameters for accurate Zeta potential determination—were measured using a liquid dielectric constant tester (GCSTD-F, Guance Ltd., China) and a Discovery Hybrid Rheometer (TA Instruments, USA), respectively. Viscosity was recorded under a constant shear rate in a double-gap Couette geometry, with five repeats per sample. Each viscosity measurement was kept as brief as possible to minimize ethanol evaporation due to its high volatility. The physical parameters of the electrolyte aqueous solutions with different concentrations are shown in Figure S1.

\section{Results and discussion}

\subsection{Ternary phase diagram}

\begin{figure}[!t]
\centering
\includegraphics[width = 0.46\textwidth]{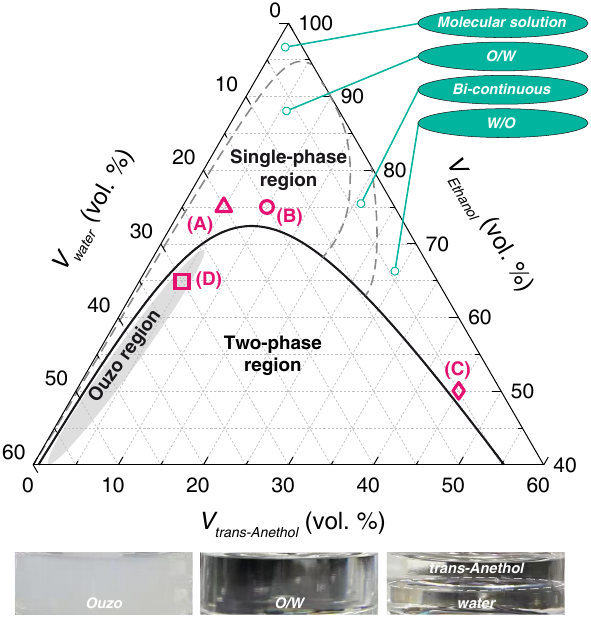}
\caption{Equilibrium phase diagram of the trans-anethol/ethanol/water ternary system investigated at 25$^{\circ}$C. The values are given in volume fraction. Four particular samples (marked in pink), with compositions close to the miscibility gap, were studied in this work. The bottom image show the macroscopic visualization of the samples in the Ouzo, O/W single-phase, and two-phase regions, respectively. } 
\label{Fig2}
\end{figure}

Figure~\ref{Fig2} illustrates the ternary phase diagram, showcasing the phase behavior of trans-anethol, ethanol as cosolvent and water at 25$^{\circ}$C. Each data point represents a sample prepared by the rapid addition of water to a trans-anethole solution in ethanol, followed by immediate vortex mixing to ensure homogeneity. The diagram delineates two primary macroscopic regions, monophasic and multiphasic, separated by a solid binodal curve. Within these regions, three distinct domains are identified based on visual observations: (i) a transparent, single-phase region at low water concentrations; (ii) a turbid, metastable region known as the Ouzo region, where spontaneous emulsification occurs due to the nucleation of trans-anethole-rich droplets as ethanol concentration decreases upon water addition; and (iii) a phase-separated region at high water concentrations, where samples demix into distinct aqueous and oil-rich layers. Visual representations of these domains are provided at the bottom of Figure~\ref{Fig2}, offering a clear depiction of the phase transitions.

The monophasic region encompasses intricate nanostructural organizations, which can be categorized into three regimes based on the internal arrangement of components: O/W nanostructures predominant at higher water contents, characterized by trans-anethole forming dispersed nanodomains within the continuous aqueous phase; bicontinuous nanostructures at intermediate compositions, featuring interpenetrating networks of oil and water domains stabilized by ethanol; and W/O nanostructures found at lower water concentrations, where water forms dispersed nanodomains within the continuous oil phase. These nanostructural regimes are delineated by dashed curves within the monophasic region of the diagram. Samples from different regions of this phase diagram were considered for subsequent analysis (marked in pink).  

\subsection{Effect of electrolyte on O/W nanostructures}

Surfactant‐free "pre‑Ouzo" microemulsions were prepared by systematically varying the NaCl concentration in the aqueous component, thereby controlling the solution’s ionic strength. In these ternary trans‑anethol/ethanol/water systems, the cosolvent (ethanol) and oil (trans‑anethol) self‐assemble into hierarchical "nanophase" structures even in the monophasic region. Previous work~\cite{li2022spontaneously} has shown that the homogeneous phase contains two coexisting populations: tiny trans‑anethol–rich molecular aggregates (of order $\sim$1 nm) and much larger mesoscopic droplets ($\sim$100 nm). In this study, the focus is on the mesoscopic droplets nucleated in the single-phase region, particularly near the miscibility gap where molecular aggregates become nearly undetectable, leaving only mesoscopic droplets in the bulk. DLS measurements were initially conducted on specific compositions (Composition-A and B) to investigate these phenomena.

\begin{figure}[!t]
\centering
\includegraphics[width = 0.46\textwidth]{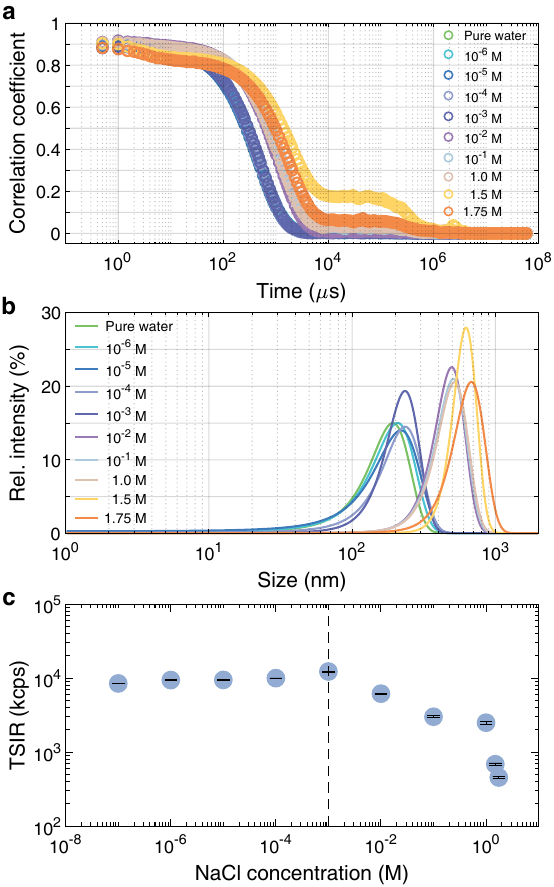}
\caption{(a) DLS intensity auto-correlation data for Composition-A with different NaCl concentrations. (b) Size distribution of the nanostructrings with increasing NaCl concentration. (c) Total scattering intensity rate (count rate, kcps) as a function of NaCl concentration.}
\label{FIG3}
\end{figure}

Figure~\ref{FIG3}(a) shows representative DLS intensity autocorrelation functions for Composition-A as NaCl concentration increases from $10^{-6}$ to 1.75 M (pure water was measured as a control). The correlation intercept (spatial coherence factor) remains high ($\sim 0.9$) under all conditions, indicating that the droplet population stays essentially monodisperse. However, at the highest salt concentrations the correlation decays slow significantly, signaling the emergence of much larger scatterers. This finding is confirmed by the size distributions in Figure~\ref{FIG3}(b): each sample shows a single well‐defined (monomodal) peak, but the peak shifts to larger diameters as NaCl is added. Specifically, the modal droplet size grows from $\sim 200$ nm at low NaCl concentration to $\sim 600$ nm at high concentration, and a fraction of droplets exceed $\sim$1 $\mu$m in diameter. Such micron‐sized droplets are comparable to classical "Ouzo" emulsions~\cite{sitnikova2005spontaneously}. In other words, increasing ionic strength does not merely broaden a fixed distribution-it drives a collective coarsening of the emulsion. The DlS results for Composition-B are shown in Figure S2, indicating a similar trend.  

Figure~\ref{FIG3}(c) plots the TSIR from DLS measurements as NaCl concentration is varied. TSIR is proportional to the sum of scattering from all nanodroplets (hence to the total volume/number of scatterers). Over a wide range (up to $\sim$$10^{-3}$ M NaCl), TSIR remains essentially constant ($\sim$$10^{4}$ kcps), implying the overall droplet population is unchanged. Above $\sim$$10^{-3}$ M, however, TSIR abruptly drops by about one order of magnitude. Because we have observed droplet sizes increasing in this regime, the sudden TSIR decrease cannot be due to shrinking droplets. Instead, it indicates a dramatic loss of scattering centers. In other words, at very high ionic strength many droplets have coalesced (or even phase‐separated) and are no longer counted by DLS, reflecting a collective destabilization of the dispersion. 

\begin{figure*}[!t]
\centering
\includegraphics[width = 0.82\textwidth]{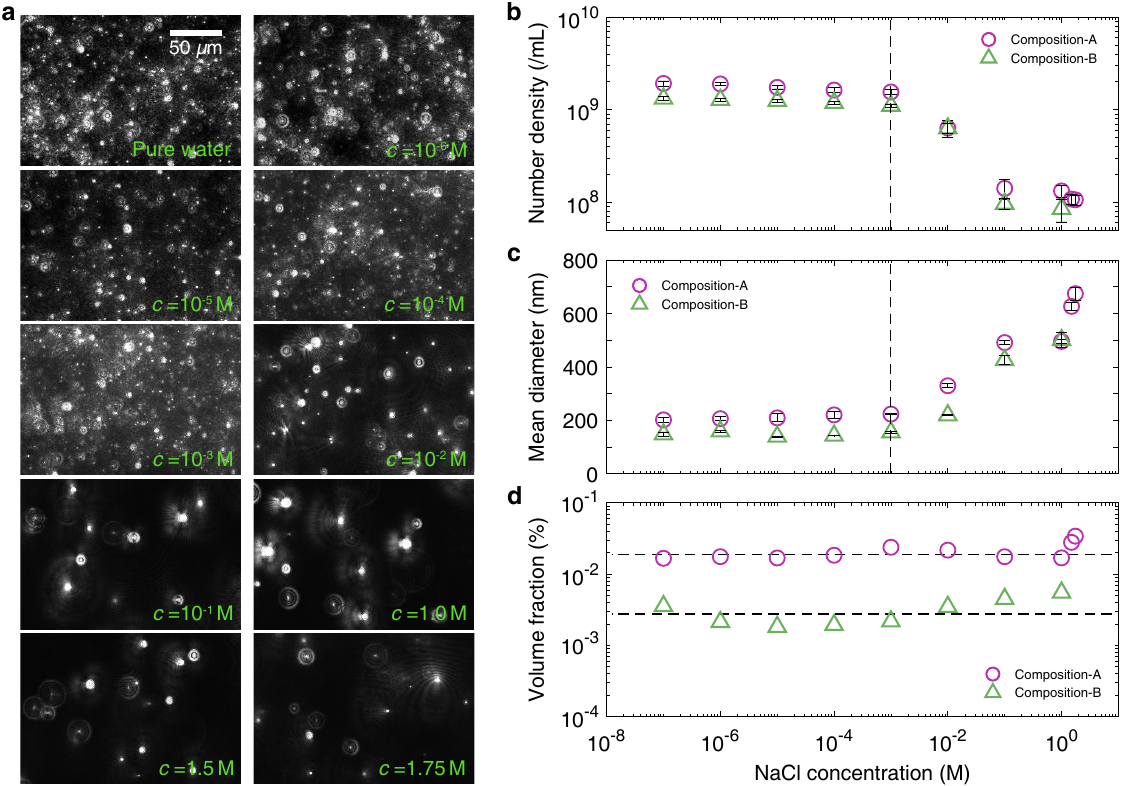}
\caption{(a) Microscopic images from the NTA experiments capturing the presence and dynamics of mesoscopic droplets for various NaCl concentrations. (b) Nucleated mesoscopic droplet concentration as a function of NaCl concentration. (c) Mean diameter of mesoscopic droplet as a function of NaCl concentration. Error bars in each panel represent the standard deviation over five measurements. (d) Effect of NaCl concentration on the volume fraction of trans-anethol component in the form of mesoscopic droplets.}
\label{FIG4}
\end{figure*}

We then employed NTA to monitor the mesoscopic oil-rich droplets by analyzing the Brownian motion of individual scatterers and extracting their size and number concentration. This approach goes beyond visual turbidity: for each salt condition we captured representative images (Figure~\ref{FIG4}(a)) and quantified the nanoparticles' number density. Notably, light scattering arises not only from the discrete trans-anethole-rich mesoscopic droplets but also from the intrinsic refractive-index (density) fluctuations in the bulk. In practice, we observed that above a critical NaCl concentration, the NTA images become dominated by the continuous phase with comparatively few bright spots, indicating a dramatic drop in droplet population. In other words, the system transitions from a pseudo-monophasic regime with many nanostructures to an effectively single-phase solution, as fewer scattering centers remain. The high image contrast at elevated salt reflects a "cleaner" continuous phase with most mesoscopic droplets suppressed.

Quantitatively, the mesoscopic droplets' number density (particles per mL) for two compositions near the miscibility boundary declines sharply once NaCl concentration exceeds $\sim$$10^{-3}$ M (Figure~\ref{FIG4}(b)). The threshold is consistent with that measured by DLS. Both samples show a marked inflection at this salt threshold, beyond which droplet counts rapidly drop. This threshold is also reflected in the measurement of the macroscopic properties of ternary solutions, e.g., viscosity (see Figure S3). By 1.0 M NaCl, nearly $90\%$ of the mesoscopic droplets have vanished. Accordingly, the average hydrodynamic diameter of the remaining nanodroplets grows substantially (from $\sim$200 nm at low salt to $\sim$600 nm at high salt). This inverse relationship between droplet number and size seems to be characteristic of electrolyte-induced coalescence or ripening.

After identifying the ionic strength dependence of nanophase-structured region, a key question is whether the "missing" mesoscopic droplets simply dissolved back into the continuous phase as molecules or ionic aggregates. To address this, we estimated the total volume of trans-anethol contained in all tracked nanodroplets (using the measured number concentration and mean diameter) and compared it to the total trans‑anethol added. Strikingly, the volume fraction of trans-anethol sequestered in mesoscopic droplets remains essentially invariant over the entire range of NaCl concentration ($\sim$$0.02\%$ for Composition-A and $\sim$$0.003\%$ for Composition-B) up to 1 M NaCl. Only at the highest NaCl level (above 1 M for Composition-A) is there a slight uptick, which we attribute to a statistical deviation when mesoscopic droplets's number density become very low. In other words, even though visible mesoscopic droplet counts fall off, the sum of droplet volumes (and thus the fraction of trans-anethol held in the dispersed phase) is conserved within experimental error. The constant volume fraction indicates that the trans-anethol simply reorganizes into fewer, larger droplets. Thus, trans-anethol is not being lost into the molecularly dissolved state; rather, it remains kinetically trapped in mesoscopic domains of roughly constant total volume. 

These observations underscore the interplay of thermodynamics and kinetics in this surfactant-free microemulsion. Thermodynamically, high salt pushes the system closer to macroscopic phase separation (as seen in the classic "Ouzo effect"), but the microscopic path takes the form of droplet coarsening rather than complete dissolution. The mesoscopic droplets persist as metastable oil-rich clusters: breaking them down into monomers would require overcoming a large interfacial free-energy barrier. In effect, the system finds a pseudo-equilibrium where a tiny but fixed fraction of oil resides in dispersed droplets. This kinetic trapping is consistent with theoretical models of charged nanodroplets, which predict Ostwald ripening behavior under reduced electrostatic stabilization. 

\subsection{Effect of electrolyte on surface charging}

\begin{figure}[!t]
\centering
\includegraphics[width = 0.46\textwidth]{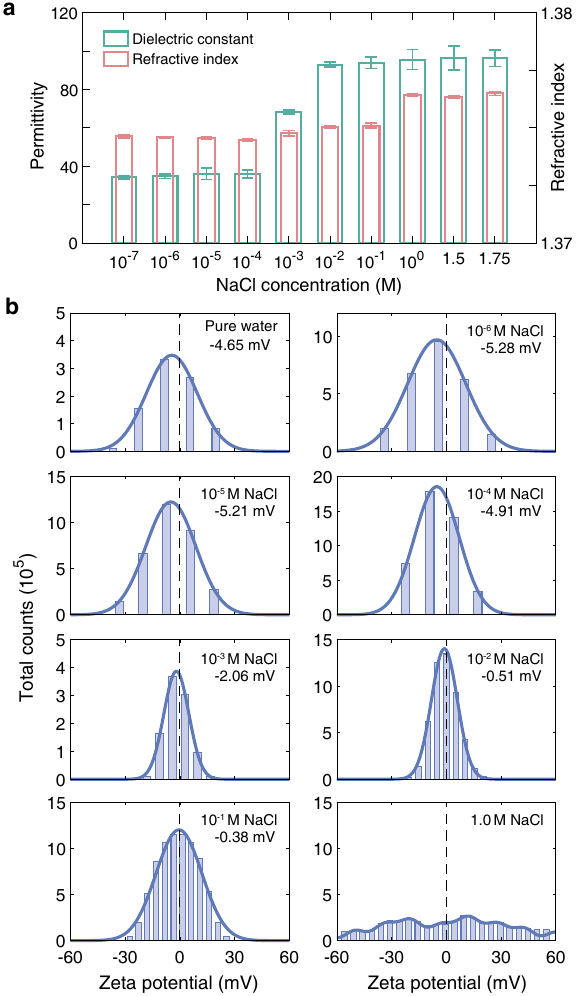}
\caption{(a) Experimental permittivity and refractive index of ternary solution as a function of NaCl concentration. The error bars denote the standard deviation over five measurements. (b) Zeta potential distribution under various NaCl concentration.}
\label{FIG5}
\end{figure}

The introduction of NaCl into the ternary trans-anethol/ethanol/water mixture adds mobile ions that profoundly modify the system's dielectric environment, molecular polarizability, and solute–solvent interactions. In particular, adding ions increases the solution's ability to screen electrostatic interactions, which affects the surface charge and stability of the mesoscopic droplets. Experimentally, we measured how the static permittivity and the refractive index of the ternary solution depend on the NaCl concentration, as shown in Figure~\ref{FIG5}(a). Both properties exhibit a pronounced transition near 10$^{-3}$ M NaCl: above this concentration, the permittivity and refractive index rise sharply with added NaCl. A higher permittivity implies that the medium can better reduce the Coulombic forces between charges and thus stabilize ionized or polar species. This enhanced polarization of the solvent also influences macroscopic properties such as viscosity and conductivity, and it can affect reaction equilibria. At very high NaCl concentrations, the permittivity approaches a saturation plateau (on the order of $\sim$100), reflecting the limit of orientational polarization of water. 

In O/W mixtures, the effective permittivity depends not only on the bulk composition but also on interfacial polarization phenomena~\cite{rambhau1992influence,wadher2020influence}. In our ternary system, as NaCl is added, the increasing ionic strength compresses the double layers and alters interfacial polarization, which contributes to the observed change in permittivity beyond what would be predicted by composition alone. By contrast, the refractive index of the solution reflects the overall microscopic polarizability and composition of the mixture. It rises if the solution contains more polarizable species or denser packing of molecules. Our measurements show that refractive index increases with NaCl concentration, indicating that the ions and their hydration shells add polarizability to the medium. The presence of ions tends to increase the refractive index because highly charged ions strongly polarize the surrounding solvent. However, this sensitivity decreases significantly as the water content decreases (see Figure S4). More generally, refractive index is influenced by solute–solvent interactions (e.g. hydrogen bonding or ion–dipole forces) that affect local density and arrangement of molecules. Therefore, measuring refractive index provides an indirect probe of the solution’s composition and the extent of ion solvation.  

To examine how added electrolytes affects droplet charge, we measured the Zeta potential of the mesoscopic droplets, as shown in Figure~\ref{FIG5}(b). We found that at low NaCl concentrations, the droplet surfaces carry a small net negative charge (on the order of -5 mV). This weak charging arises from slight differences in adsorption between anions and cations at the interface (for example, preferential adsorption of Cl$^-$ over Na$^+$). As NaCl concentration increases, the distribution of Zeta potentials shifts toward neutrality. By about 0.1 M NaCl, most droplets exhibit nearly zero Zeta potential with a narrow distribution. At even higher salt (1.0 M), no distinct positive or negative peak is resolved, indicating that the interfacial charge has been essentially neutralized. This pronounced charge depletion at high ionic strength suggests that the droplets lose their electrostatic stabilization under extreme NaCl conditions. In practical terms, high salt leads to "competitive adsorption" of Na$^+$ and Cl$^-$ and compression of the double layer, which together neutralize the interfacial charge. The reduction of surface charge lowers the electrostatic repulsion between droplets, which can promote droplet coalescence or phase separation. 

The above results (including Section 3.2) can be regarded as the consequence of multiple mechanisms. First the observed trends are consistent with classical double-layer theory and DLVO stability. Adding NaCl increases the ionic strength $I$ of the solution, which reduces the Debye screening length $\kappa^{-1}$ according to $\kappa^{-1} = [\varepsilon_r \varepsilon_0 k_B T/(2N_Ae^2 I)]^{1/2}$, where $\varepsilon_0$ is the permittivity of free space, $\varepsilon_r$ is the relative permittivity of the ternary solution, $k_B$ is the Boltzmann constant, $T$ is the temperature, $N_A$ is Avogadro’s number, $e$ is the elementary charge, and $I = \tfrac12 \sum_i c_i z_i^2$ is the ionic strength. In DLVO theory, the repulsive electrostatic potential decays as $\exp(-\kappa r)$, so higher ion concentration greatly diminishes the repulsive barrier between droplets. Droplets can approach each other more closely, facilitating coalescence. Ostwald ripening also accelerates, because the reduced double-layer forces make the system behave more like a partially miscible solution. Meanwhile, salt strongly hydrates: both Na$^+$ and Cl$^-$ bind water molecules tightly. This "salting-out" effect reduces the ability of water to solvate organic solutes (ethanol and trans-anethol). The hydrated ions effectively tie up water, lowering the effective solubility of ethanol and oil in the aqueous phase. As a result, ethanol and oil preferentially migrate into the dispersed phase or adsorb at the interface. This redistributes mass from the bulk water-rich phase into the oil-rich droplets, causing an overall shift in the droplet size distribution toward larger diameters (rightward shift rather than simple broadening).  

Molecular dynamics simulations~\cite{schottl2018salt} also provide microscopic support for this mechanism. This expulsion of ethanol from the continuous phase increases the relative polarity of the remaining water phase. Thermodynamically, one can view this as moving the mixture deeper into the two-phase region of the phase diagram. Finally, the interplay of ions with water leads to significant enthalpic and entropic effects. The strong ion–water interactions mean that adding salt lowers the chemical potential of the aqueous phase relative to the organic phase. This makes the mixed (emulsified) state less favorable compared to phase separation. At high NaCl concentration, the hydrophobic cores of the trans-anethol droplets become largely covered by ethanol molecules. These ethanol layers can sterically and electrostatically shield the oil from the aqueous ions. As a result, additional ions in solution have less impact on the interface at lower concentrations. Only when salt is very high do the ions penetrate to neutralize whatever interfacial sites remain. This nuanced interfacial adsorption, where ions preferentially localize in ethanol-poor regions of the interface~\cite{marcus2015influence}, explains why the Zeta potential changes gradually and why the stability loss is most dramatic at extreme ionic strength. 

\subsection{Effect of electrolyte on W/O reverse aggregates}

\begin{figure}[!t]
\centering
\includegraphics[width = 0.47\textwidth]{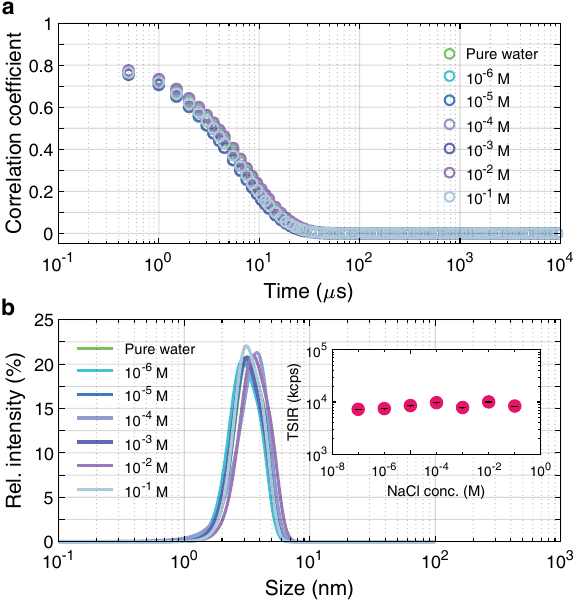}
\caption{Response of reverse aggregates (W/O nanophase) to ionic strength. (a) DLS intensity auto-correlation data for Composition-C with different NaCl concentrations. (b) Size distribution of the reverse aggregates under different NaCl concentrations. Inset: Dependence of TSIR on NaCl concentration. }
\label{FIG6}
\end{figure}

We next turned to explore the effect of the electrolyte on the reverse aggregates, i.e., W/O nanophase structurings. The DLS data for Composition-C clearly show a single‐phase, W/O nanostructure, as illustrated in Figure~\ref{FIG6}. The intensity autocorrelation functions (Figure~\ref{FIG6}(a)) and derived size distributions (Figure~\ref{FIG6}(b)) consistently indicate stable reverse aggregates roughly 1$\sim$10 nm in diameter, with a dominant peak near 3 nm. Upon increasing NaCl concentration (up to 0.1 M), the DLS profiles change only minimally: the peak position shifts by only a few angstroms at most, and the total scattering intensity remains on the order of $10^4$ kcps throughout (see the inset). In other words, the nucleation and size of the reverse (W/O) aggregates are essentially insensitive to the added electrolyte.

\begin{figure*}[!t]
\centering
\includegraphics[width = 0.97\textwidth]{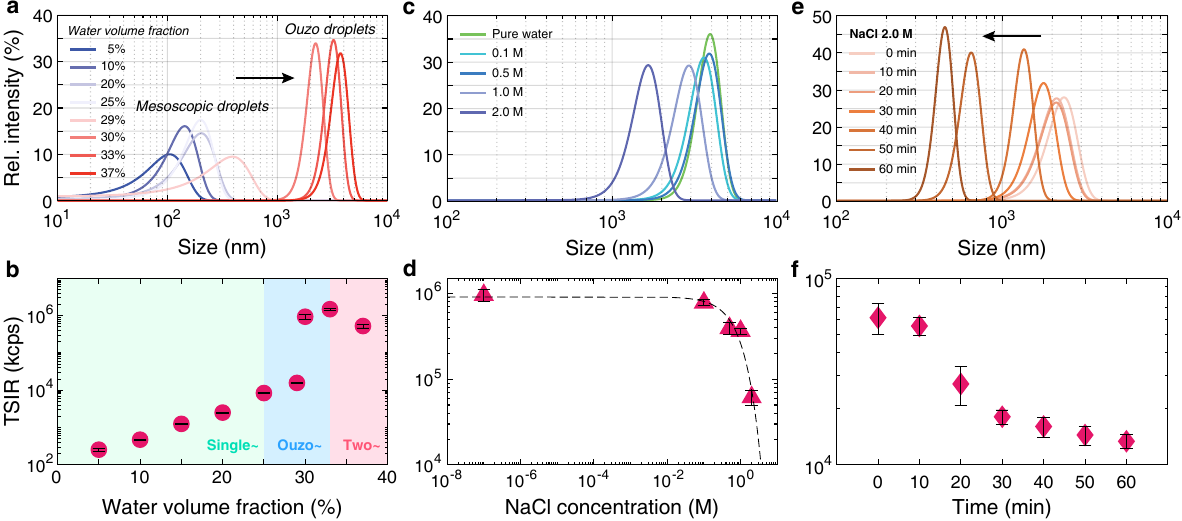}
\caption{(a) Size distribution profiles probing the continuous evolution of nano- and microstructures along a composition path from pre-Ouzo domain to the Ouzo domain as a function of the water volume fraction. (b) The TSIR of the corresponding sample in (a) as a function of water volume fraction. (c) Effect of NaCl concentration on the size distribution of nucleated Ouzo droplets. (d) The TSIR of the corresponding sample in (c) as a function of NaCl concentration. The dashed curve is the best fit. (e) Long-term monitoring the measurements of ternary solutions with 2.0 M NaCl added. (f) The TSIR of the corresponding sample in (e) as a function of time. }
\label{FIG7}
\end{figure*}

This unexpected insensitivity can be rationalized by considering the mixture’s polarity and dielectric properties. Composition-C contains trace amounts of water, so the continuous phase is predominantly an organic medium. This oil‐rich matrix has a much lower permittivity than pure water, and thus provides poor screening of electrostatic interactions. In our case the effective permittivity of the ethanol/anethol mixture is far below that of bulk water, so salt ions do not become deeply solvated in the continuous phase. Instead, they remain largely confined to the tiny aqueous cores, and in small quantity. Thus increasing ionic strength does not appreciably change the overall permittivity or polarity of the system, nor does it significantly alter the ionization state of the dispersed water-rich aggregates.

Microscopically, the robustness of these reverse aggregates is mainly governed by interfacial molecular packing and solvation forces. Short‐chain ethanol molecules act as hydrotropes at the droplet interfaces: the ethanol O-H groups form a hydrogen‐bonded shell around each water core, and the ethanol tails interact with the hydrophobic trans‐anethol phase. This results in a tightly packed interfacial layer that dramatically lowers the oil–water interfacial tension and "locks in" the aggregate structure. Finally, the mostly apolar environment of the inverse microemulsion greatly suppresses any electrostatic effects.  Indeed, theoretical calculations indicate that a typical W/O SFME droplet carries only a very small induced surface charge~\cite{han2022formation}. This near‐neutral charge state means the droplets do not strongly attract counterions, and long‐range electrostatic interactions are essentially negligible. 


\subsection{Effect of electrolyte on Ouzo emulsion droplets}

In an Ouzo mixture, the trans-anethol phase is initially soluble in the ethanol–water solution, but adding water causes microscopic oil droplets to nucleate and scatter light, making the solution appear milky. These spontaneously formed emulsions remain metastable for a long time, often much longer than common food emulsions like vinaigrette. Although a very turbid sample can impede the laser, we ensure the droplet concentration is moderate so the laser beam still penetrates the solution. In this regime DLS can reliably report the droplet size distribution.   

We first probe how the structures in the bulk evolve from the pre-Ouzo domain to the Ouzo domain. Here we first identify a compositional pathway that, by increasing the volume fraction of the water component (keeping the volume fraction of trans-anethol constant at $5\%$)), crosses the miscibility limit boundary from the single-phase O/W region and enters the Ouzo region (see Figure S5 in Supplementary material). The miscibility limit here lies at about $25\%$ water by volume. Below this limit (pre-Ouzo region), the mixture remains transparent and contains mainly nanostructured trans-anethol-rich assemblies. DLS size spectra in this region reveal droplets on the order of tens to a few hundred nanometers, as shown in Figure~\ref{FIG7}(a). Indeed, we observe a broad, low-intensity size peak up to roughly 400 nm. Once the water fraction slightly exceeds $25\%$, the solution suddenly enters the Ouzo domain. Here the tiny nanodroplets vanish from the DLS spectrum and are replaced by much larger O/W droplets. These Ouzo droplets are several micrometers in diameter (1$\sim$6 $\mu$m). This jump from submicron to micron size is a hallmark of the Ouzo effect and signals that the system has crossed into the two-phase region. In other words, adding enough water causes the trans-anethol to phase-separate in the form of large droplets, whereas in the pre-Ouzo regime the trans-anethol is only present as weakly defined nanoscale aggregates.   

Previous work~\cite{prevost2021spontaneous} has shown that metastable Ouzo emulsions often coexist with residual nano-aggregates in the bulk liquid. In our measurements within the Ouzo region, however, we did not resolve any distinct nanometer-sized modes. Most of the scattered light came from the visible Ouzo droplets. The likely explanation is that micrometer-scale droplets dominate the scattering: any weak signal from pre-Ouzo aggregates may be masked by the bright scattering of the large droplets. As we increase the water content along this path, the TSIR also changes, as shown in Figure~\ref{FIG7}(b). In our experiments, it increases continuously as the water fraction approaches about $33\%$. This rise reflects that the number or size of droplets in the sample is increasing, the system is moving toward the two-phase boundary and more trans-anethol is phase-separating into droplets. Once the water fraction exceeds $33\%$, TSIR abruptly falls. This drop signals that the sample is crossing fully into a macroscopic two-phase region: many droplets coalesce and form a continuous trans-anethol phase. In short, the TSIR peak occurs near the binodal line (phase boundary) and its decline marks the onset of large-scale phase separation.  

Next we study how adding NaCl perturbs the Ouzo emulsions. We prepared Ouzo mixtures (beyond the miscibility threshold) and added increasing concentrations of NaCl. DLS size distributions show that the peaks shift toward smaller values as NaCl is added, although most droplets are still in the $\mu$m range (Figure~\ref{FIG7}(c)). The larger droplets tend to disappear from the distribution when NaCl is present. Again, we interpret this as a classic "salting-out" effect: salt ions have a strong affinity for water and effectively reduce the solubility of the trans-anethol in the aqueous phase. Increasing ionic strength raises the oil–water interfacial tension and drives phase separation. As a result, the largest metastable droplets rapidly coalesce and drop out of the dispersed phase, leaving behind only the smaller droplets (which are more stabilized by residual ethanol). This scenario is confirmed by TSIR data: adding NaCl causes a sharp decline in TSIR (Figure~\ref{FIG7}(d)). A lower TSIR means many scatterers have vanished from the liquid, consistent with large droplets merging into a bulk phase. Note that for truly micron-sized droplets, electrical charge effects (double-layer repulsion) become negligible in comparison with gravitational and interfacial forces, so the addition of salt primarily affects solubility rather than electrostatic stability.

To further illustrate the destabilizing role of electrolytes, we followed an Ouzo sample containing 2.0 M NaCl over 60 min, as shown in Figure~\ref{FIG7}(e). Initially, the droplets had a mean diameter around 2.5 $\mu$m. Over time, the DLS size distribution continuously shifted downward: by the end of the observation period the dominant droplet size had shrunk to around 400 nm. Concurrently, TSIR fell steadily, as seen in Figure~\ref{FIG7}(f). These trends indicate that the largest droplets progressively merge into the trans-anethol-rich phase, while new nanoscale droplets nucleate in the aqueous phase. In other words, the high salt destroys the metastable Ouzo emulsion: the oil rapidly separates out, leaving behind only small residual droplets. This behavior agrees with previous reports on salting-out: the presence of a strong electrolyte increases water polarity and drives oil out of solution. Ions preferentially bind water molecules and push ethanol toward the oil interface, reducing the oil solubility in water. The net effect is an accelerated phase demixing. 

\section{Conclusions}

In this work, we systematically elucidated how added electrolyte reshapes the nanostructure and stability of surfactant-free microemulsions in a trans-anethol/ethanol/water system by combining DLS, NTA, dielectric measurements, and Zeta-potential analysis. The universal structural response has been validated, spanning the full range of solution dispersed-phase structurings in the phase diagram, from sub-10 nm W/O reverse-aggregates to O/W mesoscopic droplets ($\sim$100 nm) and classical Ouzo droplets ($\sim$1 $\mu$m). 

Generally, for the O/W nanophase structuring, we observed that adding NaCl above $10^{-3}$ M causes mesoscopic droplets to coalesce into fewer, larger entities (even up to $> 1\ \mu$m). This behavior arises from two synergistic effects: (1) Debye-layer compression reduces the DLVO repulsive barrier, enabling van der Waals attraction and accelerated Ostwald ripening; and (2) the salting-out effect, whereby hydrated $\rm {Na^+}$ and $\rm{Cl^-}$ tie up water, lowering cosolvent solubility in the aqueous phase and driving them into the interface. Notably, even as the number of mesoscopic droplets and total scattering intensity decrease sharply, the overall volume fraction of trans-anethol in suspended mesoscopic droplets remains constant roughly ($\sim$$0.01\%$). This indicates that salt ion reorganizes existing trans-anethol into larger, kinetically trapped droplets rather than dissolving them into monomers. Moreover, the NaCl in the SFMEs profoundly alters both its dielectric environment and droplet stability. As NaCl concentration exceeds the threshold, $\sim$$10^{-3}$ M, the static permittivity and refractive index rise sharply, reflecting enhanced solvent screening of Coulombic interactions and increased molecular polarizability, which is in line with previous MD simulation results~\cite{schottl2018salt}. 

However, the reverse W/O nanophase proved almost insensitive to ionic strength. In this hydrophobic continuous phase, ions cannot fully penetrate the low-dielectric medium and thus remain confined to tiny water cores. Interfacial stability is governed by tight ethanol–water hydrogen bonding and molecular packing of ethanol and trans-anethol, creating an ethanol-rich sheath that repels ions. Charge transfer between water cores and the nonpolar matrix is thermodynamically unfavorable, so ionic perturbations fail to alter droplet structure.

Third, we traced the transition from a pre-Ouzo region to a “true” Ouzo region. Once the composition system crosses the miscibility limit: nanoscale oil clusters disappear, replaced by micrometer Ouzo droplets. Further, pre-formed Ouzo droplets were destabilized by adding NaCl. Their size distribution shifts toward smaller values as large droplets coalesce and float up, nucleating smaller droplets. Long-term stability tracking experiments show that the additive salt ions destroys the metastable properties of the Ouzo system, leading to accelerated ripening and phase demixing. As a result, the metastable Ouzo emulsion collapses into a bulk oil phase and a small population of residual nanodroplets.  

These findings converge on a coherent physical-chemical picture: added NaCl compresses electrical double layers, neutralizes interfacial charge, and salts out cosolvent components, thereby shifting the balance among molecular aggregates, mesoscopic droplets, and macroscopic phase separation. In practical terms, maintaining ionic strength below $10^{-3}$ M preserves long-lived mesoscopic droplets suitable for sustained delivery or turbidity, whereas deliberate salt addition can trigger rapid coalescence or phase separation. These insights provide a blueprint for designing tunable, surfactant-free emulsions in pharmaceutical, food, and materials applications.

\section*{Declaration of Competing Interest}

The authors declare that they have no known competing financial interests or personal relationships that could have appeared to influence the work reported in this paper.

\section*{Acknowledgments}
This work is supported by National Natural Science Foundation of China under Grants Nos. 12202244 and 92252205, the Fundamental Research Project (No. 9091201), and the Oceanic Interdisciplinary Program of Shanghai Jiao Tong University (No. SL2023MS002).

\appendix
\section*{Appendix A. Supplementary Material}

Supplementary data associated with this article can be found in a separate file. 

\bibliographystyle{elsarticle-num}



\end{sloppypar}

\end{document}